\documentclass[a4paper]{jpconf}
\usepackage{graphicx}

\begin{document}

\begin{flushright}
{MAN/HEP/2007/21}
\end{flushright}

\vspace*{-1cm}

\title{Hadronic photon-photon scattering at LEP\footnote{to appear in the proceedings of HEP'07, Manchester, U.K., 19.-25. July 2007}}

\author{Thorsten Wengler}

\address{University of Manchester, School of Physics and Astronomy,
  Manchester M13 9PL, UK}

\ead{Thorsten.Wengler@manchester.ac.uk}

\begin{abstract}
Hadronic interactions of two quasi-real photons have been studied extensively
both during the LEP1 and the LEP2 data taking periods. The higher
energies available at LEP2 in particular opened regions of phase space
where hadronic processes can be predicted reliably by perturbative QCD
calculations, usually available to next-to-leading order in the strong
coupling constant for the process concerned. Over a wide range of
observables and phase space good agreement is observed between
measurements and theory, however, there are a few exceptions. The L3
collaboration has found large discrepancies for high momentum particle
and jet production between theory and experiment, and measurements of
open b-quark production by DELPHI, L3, and OPAL are consistently in
excess of the theoretical values. Three new measurements have now
become available and will be discussed in this paper: on jet and
hadron production by OPAL and open beauty production by ALEPH.
\end{abstract}

\section{Cross section for open b-quark production}

Until recently the only measurement of open b-quark production in
two-photon collisions had been published by the L3
collaboration~\cite {bib-l3-bb}. In this paper the cross section was
measured to be about three times the prediction of NLO QCD. Similar
results have been reported by DELPHI~\cite{bib-delphi-bb} and
OPAL~\cite{bib-opal-bb} at conferences. All these results have in
common that the cross section is obtained from a fit to the transverse
momentum spectrum of leptons with respect to jets.
\begin{figure}[h]
\includegraphics[width=20pc,height=14pc]{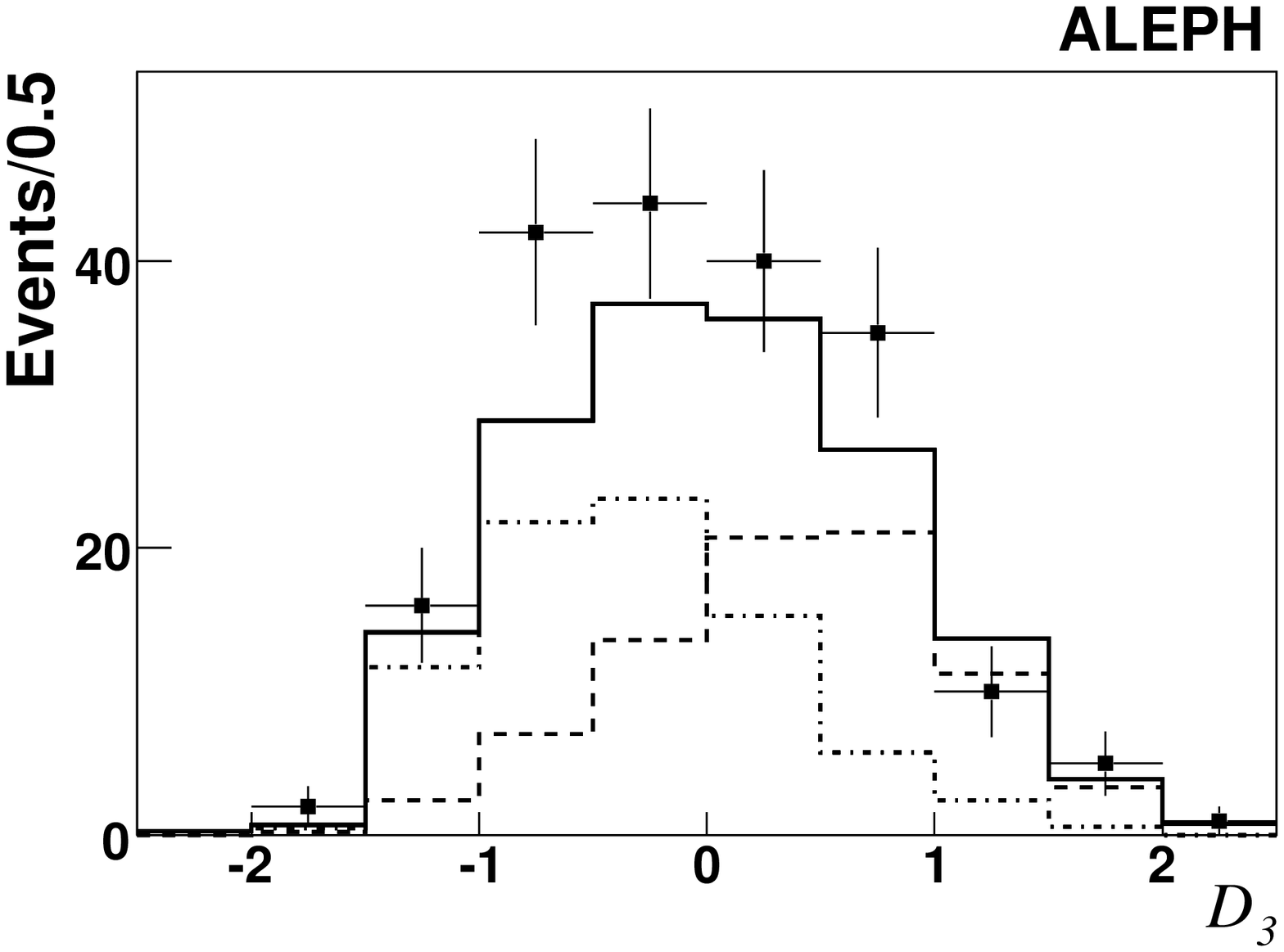}
\includegraphics[width=20pc,height=14pc]{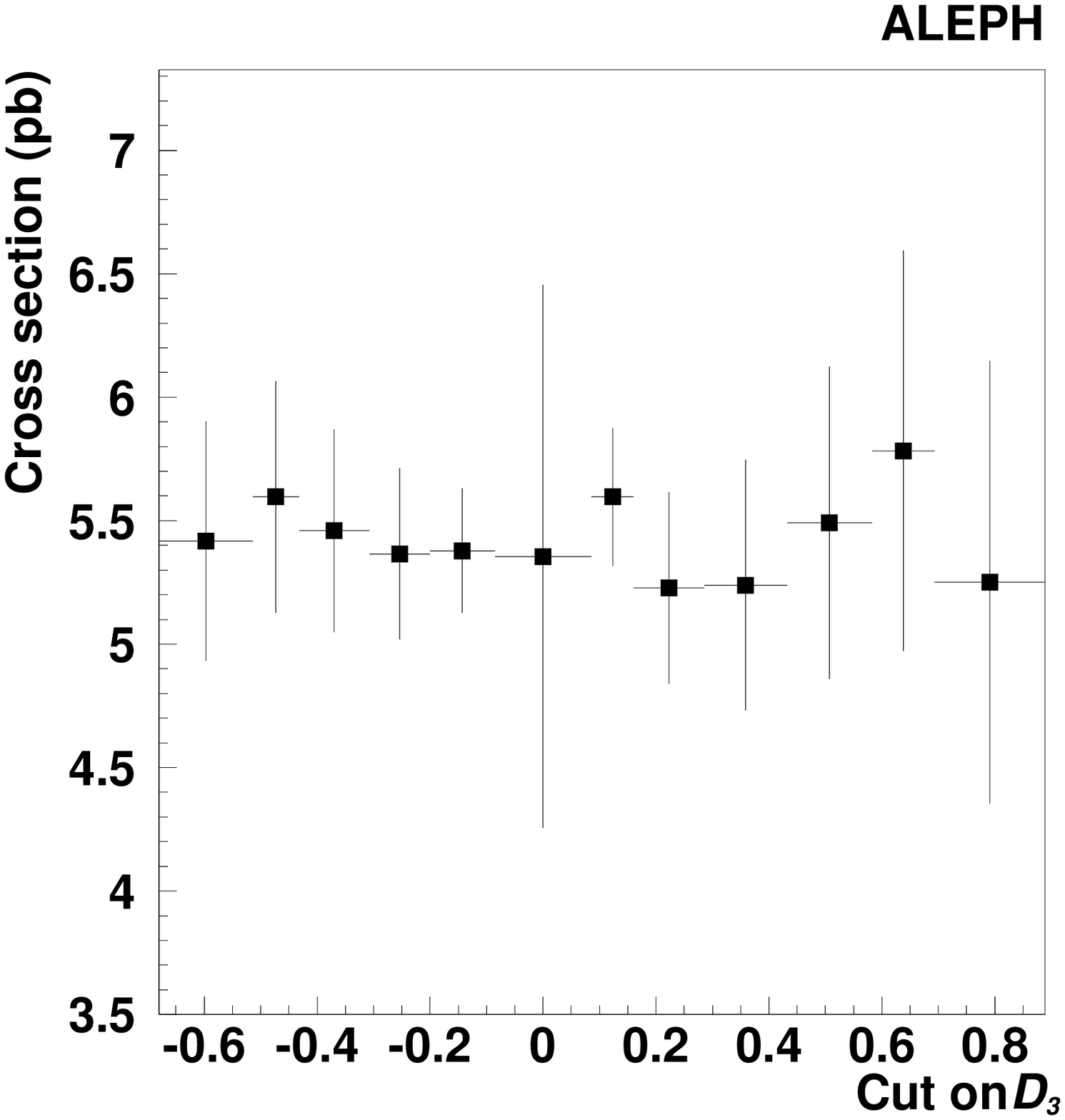}
\caption{\label{fig1}The distribution of the discriminant function for
the selection of b-quark events after the third iteration, $D_3$
(left), and the stability of the final result under variation of the
cut on $D_3$ (right), as measured by ALEPH.}
\end{figure}
This quantity is expected to be enhanced for leptons from the leptonic
decays of b-hadrons in comparison to leptons from processes involving
lighter quarks. The standard method of identifying b-decays by tagging
the displaced secondary vertex due to the longer lifetime of b-hadrons
had not been used in two-photon processes at LEP, as the lower average
momentum and hence smaller displacement of the secondary vertex made
such techniques difficult to apply. ALEPH~\cite{bib-aleph-bb} as now
published a measurement from 698 pb$^{-1}$
($\langle\sqrt{s_{\mathrm{ee}}}\rangle = 196$ GeV) using for the first
time lifetime information to tag b-quarks in two-photon interaction at
LEP. The tagging algorithm relies on the impact parameter (the
distance of closest approach to the main vertex) of tracks to identify
decay products from long lived particles. A fit to the distribution of
the ratio of the impact parameters and their estimated measurement
uncertainties is used to derive the probability that the track
originated at the main vertex. These probabilities are then combined
to probabilities that the whole event, the first jet, and the second
jet contain no decay products from long lived particles, and used as
selection criteria in the event selection. The jets are ordered such
that the mass of the first jet is closest to the b quark mass of 5
GeV/c$^2$, the mass of the second jet is the second closest,
etc. The selection is performed using Iterative Discriminant
Analysis. A total of three iterations is used, with the new
discriminant function generated for the remaining events after the cut
on the previous function has been applied. Figure~\ref{fig1} (left)
shows the discriminant function of the last iteration before the final
selection cut is applied at $D_3=0$. The right-hand side of
figure~\ref{fig1} demonstrates the stability of the resulting cross
section under variation of the final selection cut. Similar studies
have been carried out with $D_1$ and $D_2$. The final result is
$\sigma (\mathrm{e^+e^- \rightarrow e^+e^- b\bar{b}X}) = (5.4 \pm
0.8_{\mathrm{stat}} \pm 0.8_{\mathrm{syst}})$ pb, which is consistent
with the prediction of NLO QCD, but significantly lower than the
result obtained by L3: $(12.8 \pm 1.7_{\mathrm{stat}} \pm
2.3_{\mathrm{syst}})$ pb.

\section{Inclusive jet and particle production}

Another area where drastic disagreement between NLO QCD predictions
and measurement in two-photon processes has been observed is the
region of high transverse momentum ($p_\mathrm{T}$) in single jet and
particle production. The L3 collaboration has observed this excess
consistently for both charged and neutral
pions~\cite{bib-l3-pizero,bib-l3-pipm} and inclusive jet
production~\cite{bib-l3jets}. It is worth noting that earlier
measurements of these quantities will smaller data sets and hence
smaller $p_\mathrm{T}$ values found agreement with NLO
QCD~\cite{bib-opal-old-particles, bib-opal-oldjets}, as did a
measurement of di-jet cross sections using the full high energy data
of LEP2~\cite{bib-opal-dijets}.

\begin{figure}[h]
\includegraphics[width=19.5pc,height=19.5pc]{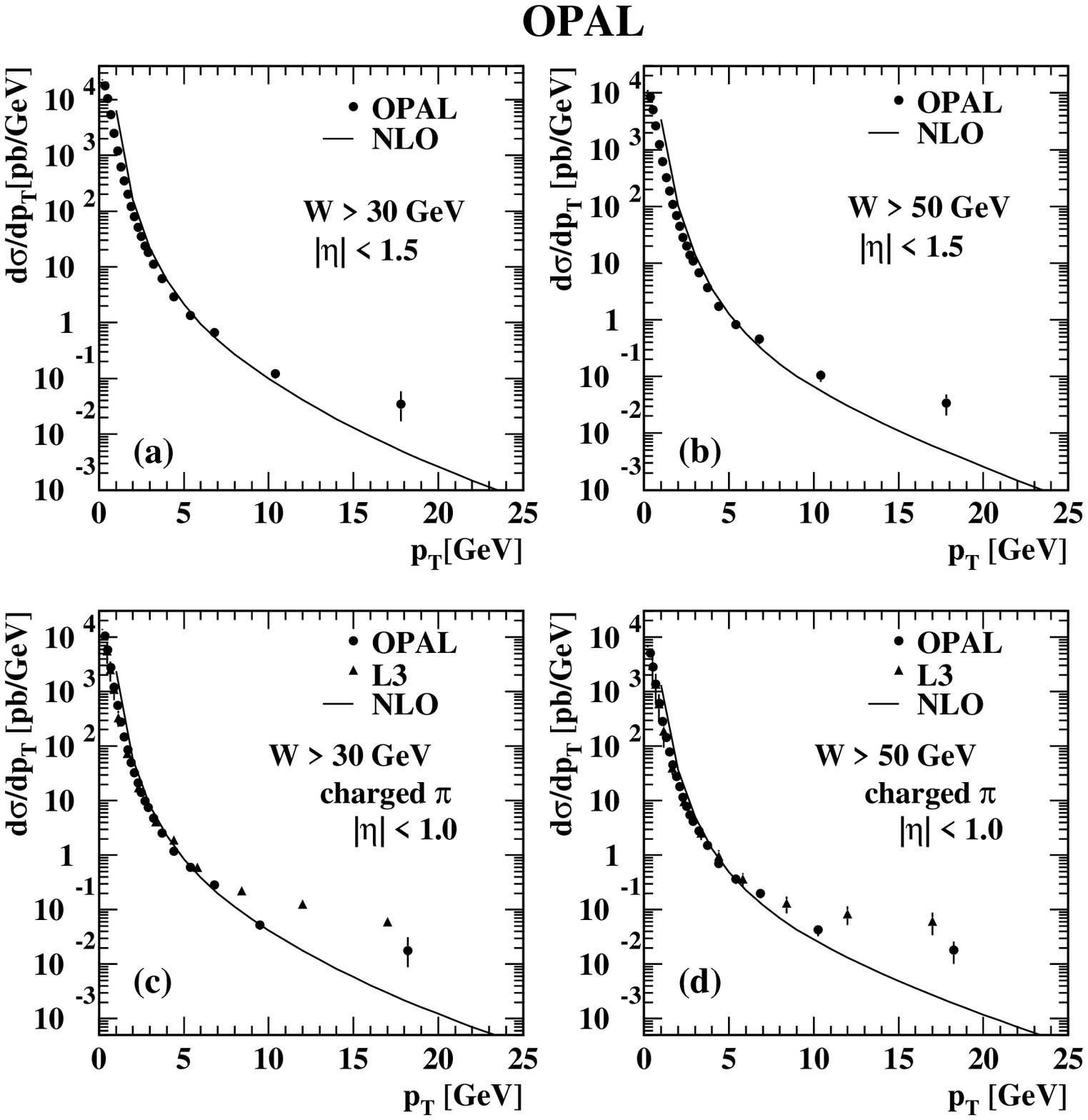}
\includegraphics[width=18pc,height=18pc]{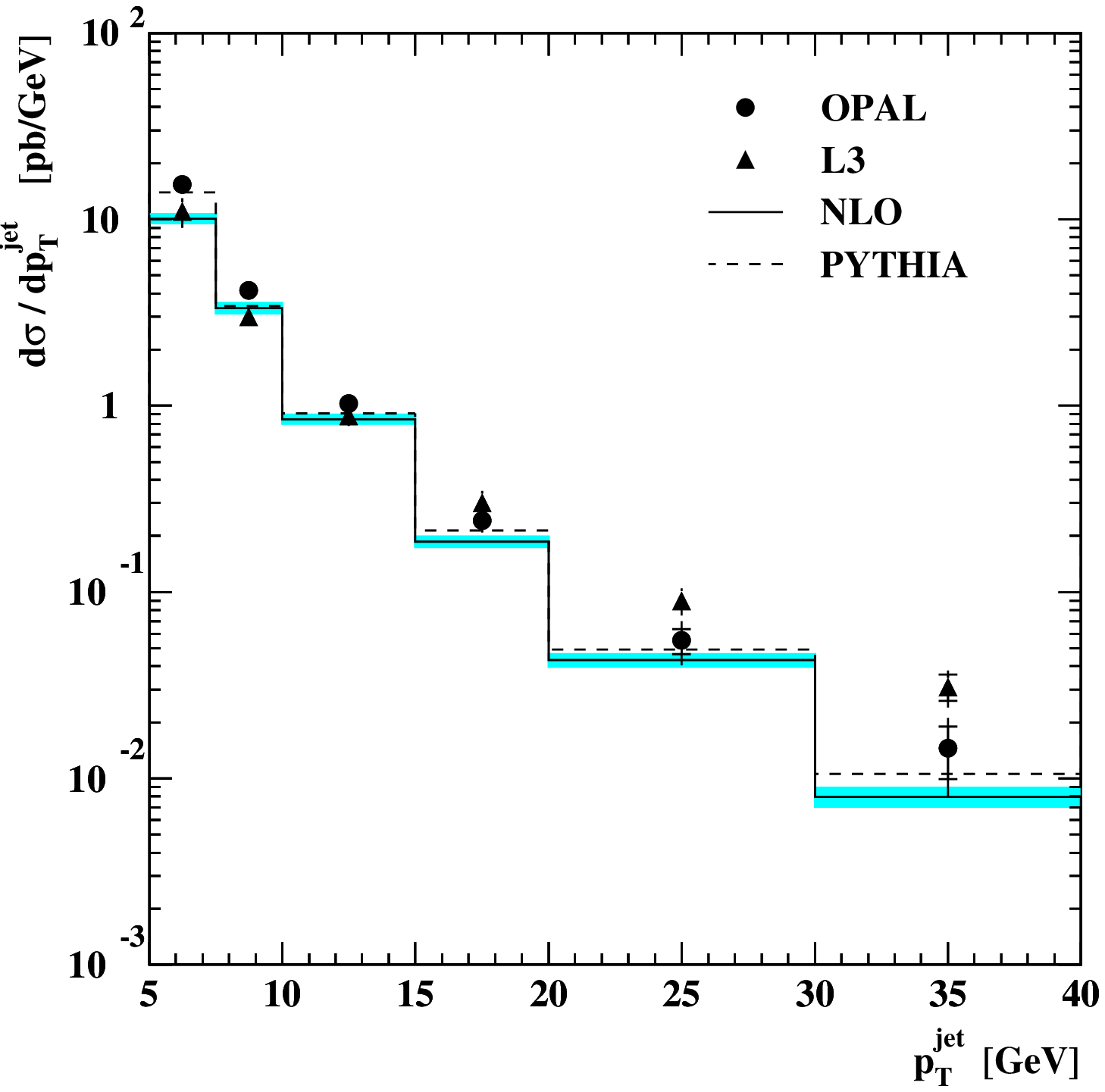}
\caption{\label{partjets}Charged pion transverse momentum spectra as
  measured by OPAL and compared to an earlier measurement by L3, and
  to NLO QCD (left), and the inclusive cross jet cross section as a
  function of the jet transverse momentum as measured by OPAL, and
  compared to an earlier L3 measurement, and the predictions of NLO
  QCD and PYTHIA.}
\end{figure}

Two new measurements have recently been published that are relevant in
this area. OPAL~\cite{bib-opal-part} has measured the production of
charged hadrons in 612.8 pb$^{-1}$
($\langle\sqrt{s_{\mathrm{ee}}}\rangle = 195.8$ GeV), in four ranges
of hadronic invariant mass, $W$, of the event. The results are
compared to an NLO QCD calculation that has been repeated for the
kinematic conditions of the measurement~\cite{bib-kniehl}. The
calculation is found to lie significantly below the data for
$p_\mathrm{T}$ greater than about 10 GeV, which can be reached
only in the highest range of 50 GeV $<W<$ 125 GeV. To facilitate a
comparison with the L3 measurement of charged
pions~\cite{bib-l3-pipm}, OPAL has repeated the measurement for $W>$
30 GeV and $W>$ 50 GeV and scaled the result for the smaller
acceptance in pseudo-rapidity of $|\eta| <$1.0 for L3 (compared to
$|\eta| <$1.5 in OPAL) and for the fraction of charged pions of all
charged hadrons using MC simulations. The results are shown on the
left of figure~\ref{partjets}. It is evident that the distributions
measured by OPAL fall more rapidly towards high $p_\mathrm{T}$
than those measured by L3. In consequence there is a disagreement
between the two experiments in this region and a better description of
the OPAL data by NLO QCD than is the case for the L3 data. The only
OPAL data point significantly higher than the calculation is that at
highest $p_\mathrm{T}$.

Sensitive to the same underlying process, but complementary in
observables and method, is the measurement of inclusive jet
production, i.e. the cross section for events with one or more jets
above a certain threshold. In case of the particle spectra discussed
above the NLO QCD calculation proceeds from the calculation of the
partonic process via fragmentation functions ~\cite{bib-afg}
determined from fits to other data sets to obtain the observable
quantities to be compared to data. Jet observables are designed to
minimize the difference between partonic and hadronic
(i.e. observable) quantities. Differences are nevertheless expected
and are studied using the string hadronisation model as implemented in
PYTHIA~\cite{bib-pythia}. OPAL~\cite{bib-opal-jets} has recently
published a measurement of inclusive jet production using 593
pb$^{-1}$ ($\langle\sqrt{s_{\mathrm{ee}}}\rangle = 198.5$ GeV) for
kinematic conditions very similar to those in the corresponding L3
publication~\cite{bib-l3jets}. Jets are reconstructed using the
$k_\perp$ jet algorithm~\cite{bib-ktclus}. OPAL has in this
measurement for the first time employed a likelihood event selection
in two-photon processes at LEP, to maximize the reach towards high
transverse momenta, where the discrepancies between data and NLO QCD
have been observed, and where background from hadronically decaying
$Z$ bosons becomes increasingly important. Nevertheless OPAL finds
that the measurement cannot be extended beyond 40 GeV in jet
transverse momentum, where the background becomes dominant. This is to
be compared to the range of up to 50 GeV in jet transverse momentum
published by L3, who find much lower values of background up to the
highest momenta studied. In contrast to the L3 results OPAL finds good
agreement between the data and NLO QCD~\cite{bib:klasen,bib:frixione}
up to the highest jet transverse momenta studied. To enable a direct
comparison with the L3 results OPAL has determined the inclusive jet
cross section originally measured for $|\eta^\mathrm{jet}|<$ 1.5 also
in the range $|\eta^\mathrm{jet}|<$ 1.0 used in the L3 paper. The
comparison of the two experimental results and with NLO QCD is shown
in figure~\ref{partjets} (right). While the data points of L3 and OPAL
are largely compatible with each other, the L3 points lie below the
OPAL points at low $p^\mathrm{jet}_\mathrm{T}$ and above at high
$p^\mathrm{jet}_\mathrm{T}$. The OPAL points are well described by the
theoretical calculation, but there is a discrepancy between theory and
the L3 results, as already observed in~\cite{bib-l3jets}. In contrast
to the results published by L3, the analysis presented by OPAL hence
finds good agreement between data and theory, and concludes that NLO
QCD is the correct theory to describe this process.

%\begin{figure}[h]
%\includegraphics[width=19pc,height=19pc]{pr420_02.eps}
%\includegraphics[width=19pc,height=19pc]{pr420_04.eps}
%\caption{\label{label}Figure caption for first of two sided figures.}
%\end{figure}

\end{document}